\documentclass[elsarticle,twocolumn,showpacs,floatfix,nobibnotes,superscriptaddress,longbibliography,nofootnotesinbib]{revtex4-1}

\usepackage{amsmath}
\usepackage{inputenc}
\usepackage{amssymb}
\usepackage{graphicx}
\usepackage{verbatim}
\usepackage{epsfig}
\usepackage{bm}
\usepackage{color}
\usepackage{float}
\usepackage{dcolumn}
\usepackage{tabularx}
\usepackage{multirow} 
\usepackage{booktabs}
\usepackage[unicode=true,
linktocpage,
linkbordercolor={0.5 0.5 1},
citebordercolor={0.5 1 0.5},
colorlinks=true,
linkcolor=blue,
citecolor=blue,
urlcolor=blue]{hyperref} 
\usepackage{lipsum} 
\usepackage{placeins} 

\usepackage[normalem]{ulem}

\graphicspath{{.}}

\begin{document}
	
\title{Gaussian process error modeling for chiral effective-field-theory calculations of $np\leftrightarrow d\gamma$ at low energies}
	
\author{B.~Acharya}
\affiliation{Institut f\"ur Kernphysik and PRISMA$^+$ Cluster of Excellence, Johannes Gutenberg-Universit\"at Mainz, 55128 Mainz, Germany}
	
\author{S.~Bacca}
\affiliation{Institut f\"ur Kernphysik and PRISMA$^+$ Cluster of Excellence, Johannes Gutenberg-Universit\"at Mainz, 55128 Mainz, Germany}
\affiliation{Helmholtz-Institut Mainz, Johannes Gutenberg-Universit\"at Mainz, D-55099 Mainz, Germany}

\date{\today}

\begin{abstract}

We calculate the energy-dependent cross section of the $np\leftrightarrow d\gamma$ process in chiral effective field theory and apply state-of-the-art tools for quantification of theory uncertainty. We focus on the low-energy regime, where the magnetic dipole and the electric dipole transitions cross over, including the range relevant for big-bang nucleosynthesis. Working with the leading one- and two-body electromagnetic currents, we study the order-by-order convergence of this observable in the chiral expansion of the nuclear potential. We find that the Gaussian process error model describes the observed convergence very well, allowing us to present Bayesian credible intervals for the truncation error with correlations between the cross sections at different energies taken into account. We obtain a 1$\sigma$ estimate of about 0.2\% for the uncertainty from the truncation of the nuclear potential. This is an important step towards calculations with statistically interpretable uncertainties for astrophysical reactions involving light nuclei. 
	
\end{abstract}


\maketitle

Over the past couple of decades, effective field theories (EFTs) have emerged as powerful tools to study nuclear structure and reactions within systematic frameworks where a well-justified quantification of theory uncertainties is feasible~\cite{Epelbaum:2008ga}. An EFT is a model-independent description of interacting particles at some low-momentum scale $Q$ in terms of the minimal degrees of freedom relevant below a breakdown momentum scale, $\Lambda$. Physical observables are calculated as systematic expansions in $Q/\Lambda$, with 
undetermined parameters called low-energy constants (LECs).
The EFT can predict an observable at a given order once all the LECs that appear up that order have been fixed, \emph{e.g.,} by fitting to experimental data for other observables. The truncation of the EFT expansion at a finite order results in a theory error that can be estimated. Recent adoption of Bayesian methods, which have found increasing prominence in nuclear theory~\cite{Schindler:2008fh,Furnstahl:2014xsa,Furnstahl:2015rha,Wesolowski:2015fqa,Zhang:2015ajn,Melendez:2017phj,Wesolowski:2018lzj,Lovell:2018bao,Neufcourt:2018syo,Zhou:2018bon,King:2019sax,Melendez:2019izc,Premarathna:2019tup,Zhang:2019odg,Epelbaum:2019zqc,Phillips:2020dmw,Volkotrub:2020lsr,Reinert:2020mcu,Wesolowski:2021cni,Rijneveen:2020qbc,Maris:2020qne,Wellenhofer:2021eis,Connell:2021qcd,Rijneveen:2021bfw}, have enabled a statistically meaningful estimation of truncation errors which can be easily combined with other uncertainties that arise from experimental data, calculational methods, and parameter fitting (see, \emph{e.g.,} Ref.~\cite{Wesolowski:2021cni}). Such a complete and consistent accounting of theory uncertainties is of paramount importance not only when the theoretical results have to be compared to precise experiments, but also when theoretical predictions need to be used as proxies for experimental data that are imprecise or unavailable. The latter situation is common for nuclear reactions in astrophysics and cosmology, for example in those relevant to big-bang nucleosynthesis (BBN).

There is, in general, a good agreement between simulations of nuclear reaction networks and astronomical observations for abundances of light nuclei that were produced during BBN. This agreement stands not only as a historic cornerstone of big-bang cosmology, but also as a crucial test for future extensions to the Standard Model. The ability of BBN to constrain key quantities such as cosmic baryon density rests heavily on the primordial Deuterium to Hydrogen abundance ratio, whose uncertainty is currently dominated by the uncertainties of the Deuterium-burning processes: ${}^2$H$(p,\gamma){}^3$He, ${}^2$H$(d,n){}^3$He and  ${}^2$H$(d,p){}^3$H. Improvements in precision on their rates, exemplified by the recent 3\% measurement  of  the ${}^2$H$(p,\gamma){}^3$He cross  section  by  the  LUNA  collaboration~\cite{Mossa:2020gjc}, will eventually call for an update of the primordial Deuterium production reaction, $np\rightarrow d\gamma$. This reaction is the first step in the BBN network and marks the end of the so-called Deuterium bottleneck. BBN simulations~\cite{Serpico_2004,Pisanti:2007hk,Coc:2019rza,Pitrou:2018cgg} rely on theoretical predictions for this important cross section because available experimental data are too sparse and not sufficiently precise in the relevant energy regime. It is therefore imperative to study it using different theoretical approaches and to rigorously quantify uncertainties. Furthermore, ever increasing precision of the constraints on the primordial Deuterium abundance from cosmic microwave background analysis~\cite{Planck:2018vyg} and observational astronomy of low-metallicity stars~\cite{Cooke:2017cwo} need to be independently matched by nuclear physics determinations.

This reaction has traditionally been studied using phenomenological models for nuclear interactions and currents~\cite{Arenhovel:1990yg,Carlson:1997qn,Marcucci:2004sq}.  Pionless EFT, the low-energy EFT of quantum chromodynamics (QCD) that uses nucleons as the only dynamical degrees of freedom and has a breakdown scale $\Lambda$ at the pion mass $m_\pi$, was applied to $np\rightarrow d\gamma$ at BBN energies in Refs.~\cite{chen-savage,rupak,ando}. By fitting an LEC in the $E1$ amplitude to photodissociation data and an LEC in the $M1$ amplitude to $p(n,\gamma)d$ measurement~\cite{cokinos-melkonian} at energy $E=1.2625 \times 10^{-8}~\mathrm{MeV}$ in the $np$ center-of-mass frame, Ref.~\cite{rupak} quoted a sub-percentage precision for the cross section at BBN energies. The pionless EFT results of Refs.~\cite{rupak,ando}, along with the accompanying uncertainties, are adopted by modern BBN simulations~\cite{Pisanti:2007hk,Pitrou:2018cgg}.

This cross section has also been calculated at $E=1.2625 \times 10^{-8}~\mathrm{MeV}$ in Ref.~\cite{Beane:2015yha} using lattice QCD, in Ref.~\cite{piarulli} using chiral EFT ($\chi$EFT) and in Ref.~\cite{song-lazauskas-park} using a hybrid approach that utilized $\chi$EFT currents and phenomenological potentials. Based on the scale hierarchy $Q\sim m_\pi<\Lambda$, $\chi$EFT provides a description of hadronic and nuclear phenomena in terms of nucleons and pions as effective degrees of freedom, and is therefore well suited for investigating this cross section at energies beyond the thermal capture, including the cosmological  [20~keV--200~keV] range~\cite{PhysRevLett.82.4176} and higher.

In this work, we present the first $\chi$EFT results for the energy-dependent $np\leftrightarrow d\gamma$ cross section for a range of energies from threshold to MeV scale, that encompasses the BBN regime. We extend the Bayesian procedure for estimating the $\chi$EFT truncation error, which was earlier applied to various other observables---nucleon-mass expansion~\cite{Schindler:2008fh,Wesolowski:2015fqa}, nucleon-nucleon elastic scattering~\cite{Furnstahl:2015rha,Melendez:2017phj,Wesolowski:2018lzj,Melendez:2019izc,Reinert:2020mcu}, nucleon-deuteron scattering~\cite{Epelbaum:2019zqc,Volkotrub:2020lsr}, nuclear-matter equation of state~\cite{PhysRevC.102.054315,PhysRevLett.125.202702}, pion-photoproduction on the nucleon~\cite{Rijneveen:2020qbc,Rijneveen:2021bfw}, and properties of light nuclei~\cite{Maris:2020qne,Wesolowski:2021cni}---to an  electromagnetic reaction cross section. We adopt the error model developed by Melendez \emph{et al.}~\cite{Melendez:2019izc} that  used machine learning with Gaussian processes (GPs) calibrated by physics-based hyperparameters to determine the convergence of EFT predictions that may be correlated across independent variables (\emph{e.g.,} energy). It allows us to quantify, for the first time, the error in the $np\rightarrow d\gamma$ cross section,  $\sigma_{np}(E)$, and in that of the reverse process, $\sigma_{\gamma d}(E)$, from truncation of the $\chi$EFT potential as a Bayesian degree-of-belief band.

We begin by presenting $\chi$EFT results for the $np\leftrightarrow d\gamma$ cross section and related observables in Sec.~\ref{sec:npdagmma}. We then briefly introduce the Gaussian process error model in Sec.~\ref{sec:gpintro}. The fitted model, diagnostic checks and predicted uncertainties are presented in Sec.~\ref{sec:gpmodel}. We close with a brief summary and outlook in Sec.~\ref{sec:conclusion}.


\section{The $np\leftrightarrow d\gamma$ cross section in $\chi$EFT}
\label{sec:npdagmma}

The detailed-balance principle relates the $np\rightarrow d\gamma$ cross section $\sigma_{np}$ to the deuteron photodissociation $d\gamma \rightarrow np$  cross section $\sigma_{\gamma d}$:
\begin{equation}
    \sigma_{np} = \frac{3}{2}\,\frac{(s-m_d^2)^2}{(s-4m^2)(s-\delta m^2)}\,\sigma_{\gamma d}\,,
\end{equation}
where $m_d$ is the deuteron mass, $m$ is the isospin-averaged nucleon mass, and $\delta m$ is the neutron-proton mass difference. The Mandelstam variable $s$ can be conveniently expressed in terms of the $np$ relative energy $E$, or the neutron energy $E_n$ in the rest frame of the proton,
\begin{equation}
    s = ( E + 2 m )^2 = 2 m_p E_n + 4 m^2\,,
\end{equation}
where $m_p$ is the proton mass. 
The photodissociation cross section can be expressed as a function of the photon energy $\nu$ (in the rest frame of the deuteron) through  
\begin{equation}
\sigma_{\gamma d}(\nu) = \frac{2\pi^2}{\nu} \, \alpha \, R_T(\nu,\nu)
\end{equation} in terms of the transverse response function of the deuteron, defined as
\begin{align}
\label{eq:r_t}
  R_T(\nu,q) =  \frac{1}{6} \sum_{M} \sum_{S^\prime S_z^\prime} & \sum_{T^\prime} 
  \int\frac{\mathrm{d}^3k}{(2\pi)^3} 
                  \, \delta(\nu+m_d-E_+-E_-) \nonumber\\
                  & \sum_{\lambda=\pm1} \vert  \langle\phi_{\mathbf{k},S^\prime S_z^\prime,T^\prime}\vert j_\lambda\vert\psi_{M}\rangle \vert^2\,.
\end{align}
Here, $j_\lambda$ are the spherical components of the electromagnetic current operator ${\bf j}$ at four-momentum transfer $(\nu,\mathbf{q})$. The deuteron ground state is denoted by $| \psi_{M}\rangle$,  where $M$ is the projection of the total angular momentum, whereas 
$|\phi_{\mathbf{k},S^\prime S_z^\prime,T^\prime}\rangle $ denotes the $pn$ scattering state with the relative momentum, total isospin, total spin and spin projection given by $\mathbf{k}$, $T^\prime$, $S^\prime$ and  $S_z^\prime$ respectively. $E_\pm$ are the energies of the final-state nucleons in the rest frame of the deuteron, \emph{i.e.,} $E_\pm = \sqrt{(\mathbf{q}/2 \pm \mathbf{k})^2+m^2}$. The deuteron and the $np$ scattering-state wave functions are obtained by solving their Lippmann-Schwinger equations using the momentum-space regularized semilocal potentials of Ref.~\cite{Reinert:2017usi} at various orders in the $\chi$EFT expansion. The electromagnetic currents were first derived in $\chi$EFT in Refs.~\cite{Park:1995pn,Pastore:2008ui,Kolling:2009iq}. We use multipole expansions~\cite{Bijaya_neutrino,chi-disp-mu-d} of these  operators~\cite{bibnote1} that contribute at orders $(Q/\Lambda)^{-3,-2,1,0}$, namely the one-body currents consisting of the convection and spin-magnetization terms, and the leading one-pion exchange currents. Although we retain the quadrupole and octupole operators in the multipole expansions, their contributions are negligible compared to the dominant $M1$ and $E1$ operators at low energies. To compare the relative strengths of the $M1$ and $E1$ transitions, we calculate the analyzing power, defined as $\Sigma(\theta)\equiv(N_\parallel-N_\perp)/(N_\parallel+N_\perp)$, where $N_\parallel (N_\perp)$ is the number of outgoing neutrons parallel (perpendicular) to the photon-polarization plane in a photodissociation experiment. It is related to the $M1$ and $E1$ contributions to the photodissociation cross section through~\cite{ando}
\begin{equation}
   \Sigma(\theta) = \frac{\frac{3}{2}\sigma_{\gamma d}^{E1}\sin^2\theta}{\sigma_{\gamma d}^{M1}+\frac{3}{2}\sigma_{\gamma d}^{E1}\sin^2\theta} \,,
\end{equation}
where $\theta$ is the angle of the neutrons with respect to the photon beam axis. 

We now present the $\chi$EFT predictions for observables related to $np\leftrightarrow d\gamma$ starting with the $p(n,\gamma)d$ cross section for thermal neutrons in Table~\ref{tab:threshold}. Here (and throughout this work), we denote cross sections calculated with an $n$th-order (N$^n$LO) potential of Ref.~\cite{Reinert:2017usi} (and electromagnetic currents fixed to one-body plus two-body one-pion exchange) as $y_n$. The contribution $y_1$ is zero because there are no corrections to the $\chi$EFT potential at this order.  At this threshold energy, our results undershoot the experimental value of Ref.~\cite{cokinos-melkonian} by a few percent. It was shown in Ref.~\cite{piarulli}, that percentage-level agreement with the experiment of Ref.~\cite{cokinos-melkonian} required current operators at the order $(Q/\Lambda)^1$ not considered in this work; these included leading two-pion exchange, sub-leading corrections to one-pion exchange and contact operators. Since the goal of this paper is to apply machine learning tools for the first time to uncertainty quantification in nuclear electroweak reactions in $\chi$EFT and including these operators significantly complicates the calculations, we work with fixed currents and focus on analyzing convergence in the $\chi$EFT potential only, leaving the inclusion of higher order currents to future work. We discuss the implications of our findings for such a study further below.

\begin{center}
\begin{table}[htbp]
\begin{tabularx}{0.75\columnwidth}{cccccc}
\hline
~$y_0$~ & ~$y_2$~ & ~$y_3$~ & ~$y_4$~ & ~$y_5$~ & Experiment\\
\hline
340.6~ & 325.0~ & 321.8~ & 321.0~& 322.6~ & 332.6 $\pm$ 0.7 \\
\hline
\end{tabularx}
\caption{ The $p(n,\gamma)d$ cross section $\sigma_{np}$ in mb for thermal neutrons (corresponding to $E=1.2625\times10^{-08}$~MeV) calculated using $\chi$EFT potentials at various orders. The experimental result is from Ref.~\cite{cokinos-melkonian}.
\label{tab:threshold}}
\end{table}
\end{center}

\begin{figure*}[th]
   	\centering
   	\includegraphics[width=0.8\textwidth]{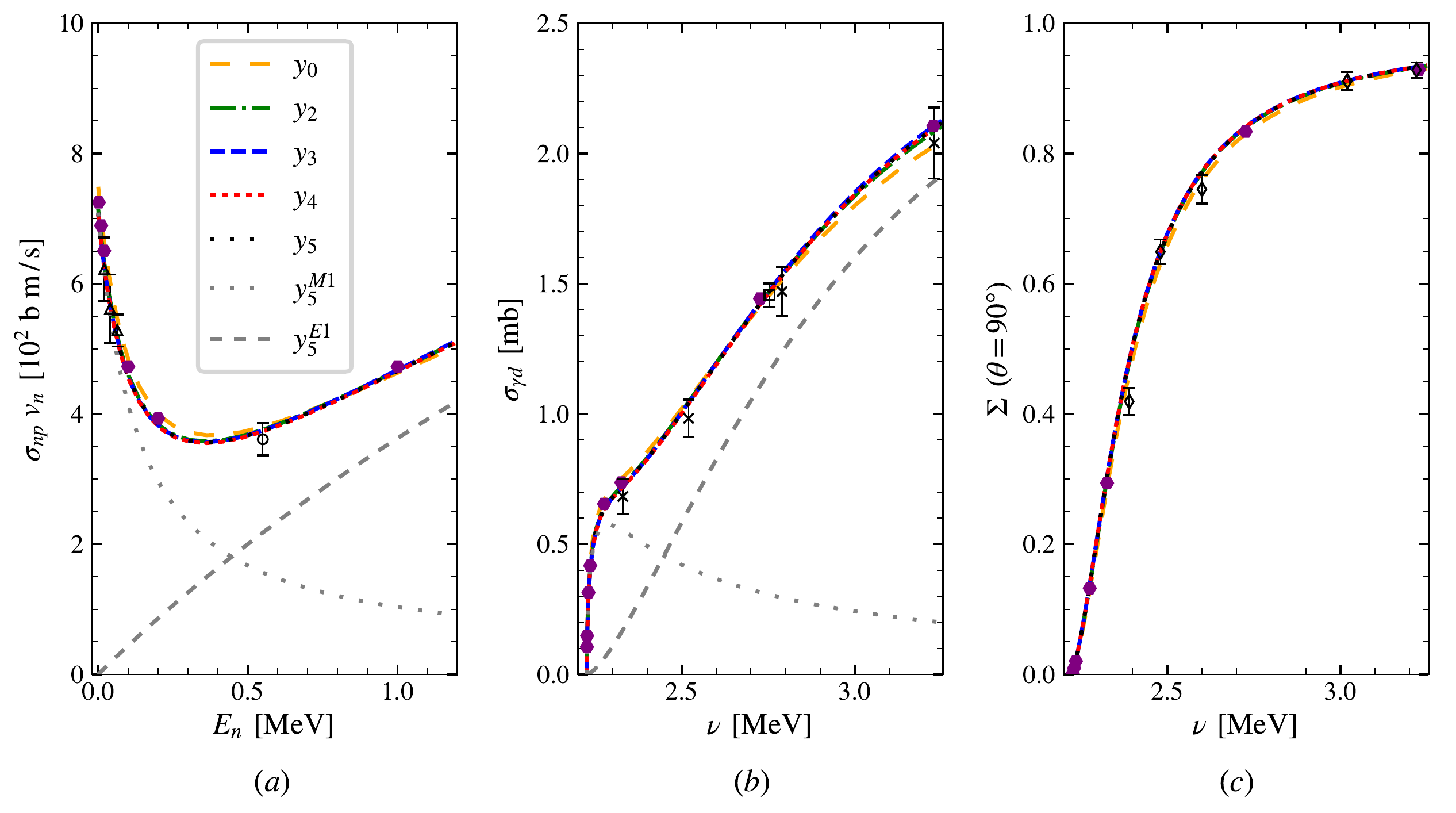}
   	\caption{$\chi$EFT predictions for the $np\leftrightarrow d\gamma$ cross section and related observables at BBN energies. (a) The product of $p(n,\gamma)d$ cross section and the neutron speed versus the energy of the neutron. (b) The deuteron photodissociation cross section as a function of the photon energy in the rest frame of the deuteron. (c) The photon analyzing power for photodissociation versus its energy. Experimental data are from Refs.~\cite{suzuki}~(triangles), \cite{nagai}~(circle), \cite{hara}~(crosses), \cite{moreh}~(square) and \cite{tornow}~(diamonds). Beam-energy resolution errors are not shown. Purple hexagons are calculated using pionless EFT results of Ref.~\cite{rupak}. 
   		}
   	\label{fig:compare_to_expt}
\end{figure*}

In Fig.~\ref{fig:compare_to_expt}, we compare our $\chi$EFT predictions at different orders for various observables related to $np\leftrightarrow d\gamma$ with the pionless EFT results of Ref.~\cite{rupak} and available experimental data at energies relevant for BBN and beyond. The $\chi$EFT predictions agree with the experimental data for all orders. Since Ref.~\cite{rupak} uses the experimental cross section given in Table~\ref{tab:threshold} as input, their $M1$ contribution is larger than our $\chi$EFT predictions at low energies.

\section{Gaussian process model for correlated EFT errors} 
\label{sec:gpintro}

Following the formalism introduced by Melendez \emph{et al.}~\cite{Melendez:2019izc}, we consider an observable $y$ as a function of the kinematic variable $p$, which, in our case is the $np$ relative momentum. The order-$k$ EFT prediction is written as 
\begin{equation}
    y_k(p) = y_\mathrm{ref}(p) \sum_{n=0}^k c_n(p) \, [Q(p)/\Lambda]^n\,,
\end{equation}
and its EFT truncation error as 
\begin{equation}
\label{eq:trunc_err}
    \delta y_k(p) = y_\mathrm{ref}(p) \sum_{n=k+1}^\infty c_n(p) \, [Q(p)/\Lambda]^n\,.
\end{equation}
Here $y_\mathrm{ref}(p)$ is a dimensionful quantity that sets the overall scale such that the \emph{a priori} unknown dimensionless coefficients $c_n(p)$ are smooth naturally-sized (\emph{i.e.,} of order 1) curves, provided that the EFT is converging in the expected manner. The Bayesian model for EFT error quantification seeks to build upon this prior knowledge with the data, the values of $c_{n\leq k}(p)$ evaluated from order-by-order EFT calculations up through order $k$, to refine our expectations for $c_{n>k}(p)$, and thereby obtain an estimate for $\delta y_k(p)$. Refs.~\cite{Furnstahl:2014xsa,Furnstahl:2015rha,Wesolowski:2015fqa,Melendez:2017phj,Wesolowski:2018lzj,Melendez:2019izc} developed a pointwise error model that is applicable when (i) $y$ is a number and not a function of one or more kinematic variables; (ii) $y(p)$ is analyzed at only one value of $p$ or at multiple values of $p$ that are sufficiently far apart such that the $y(p)$ values can be safely assumed to be uncorrelated. Ref.~\cite{Melendez:2019izc} extended this framework to study curvewise convergence by employing GPs. Below we first introduce the GP error model and then present application to $np\leftrightarrow d\gamma$.

\paragraph*{The GP error model---}The basic idea of the error model is to use GPs to build a stochastic representation (an \emph{emulator}) to serve as a surrogate for the sequence of deterministic calculations (the \emph{simulator}) that yields order-by-order EFT predictions up through order $k$. The statistical properties of the emulator are then exploited to yield estimates for the EFT truncation errors at various orders, subsuming the contributions of even those terms that have never been calculated. Specifically, we start with the assumption that $c_n(p)$ are independent draws from an underlying GP, \emph{i.e.,} they follow a multivariate normal distribution for every finite set of $p$. The GP is completely specified by the mean $\mu$ and covariance function  $\bar{c}^2r(p,p^\prime;\ell)$, 
\begin{equation}
\label{eq:cpgp}
c_n(p) \,\vert\, \bar{c}^2, \ell \, \stackrel{\mathrm{iid}}{\sim} \, \mathcal{GP}[\mu,\bar{c}^2r(p,p^\prime;\ell)]\,.
\end{equation}
The correlation function $r(p,p^\prime;\ell)$ is commonly chosen to have a squared-exponential form, 
\begin{equation}
    r(p,p^\prime;\ell) = \exp\left[-\frac{(p-p^\prime)^2}{2\ell^2}\right]\,. 
\end{equation} The correlation length $\ell$, the mean $\mu$ and the marginal variance $\bar{c}^2$ are the hyperparameters of the GP, which are learned from the training data set that comprises order-by-order EFT calculations up through order $k$. Remarkably, it follows from Eq.~\eqref{eq:cpgp} that the truncation error defined by  Eq.~\eqref{eq:trunc_err} has the distribution 
\begin{equation}
\label{eq:trunc_err_distr}
    \delta y_k(p) \,\vert\, \bar{c}^2, \ell, Q(p), \Lambda \sim 
    \mathcal{GP}[m_{\delta k}(p),\bar{c}^2R_{\delta k}(p,p^\prime;\ell)]\,,
\end{equation}
where
\begin{equation}
\label{eq:mean_trunc_err}
  m_{\delta k}(p) = \frac{y_\mathrm{ref}(p)}{\Lambda^k}\frac{Q(p)^{k+1}}{\Lambda-Q(p)}\,\mu\,,
\end{equation}
and 
\begin{align}
\label{eq:corr_trunc_err}
  R_{\delta k}(p,p^\prime;\ell) = \frac{y_\mathrm{ref}(p)y_\mathrm{ref}(p^\prime)}{\Lambda^{2k}}\frac{[Q(p)Q(p^\prime)]^{k+1}}{\Lambda^2-Q(p)Q(p^\prime)}\,r(p,p^\prime;\ell)\,.
\end{align}
With point estimates for $\ell$ and $Q(p)$, the normal-inverse-$\chi^2$ prior serves as the conjugate prior for the hyperparameters $(\mu,\bar{c}^2)$ of the Gaussian processes above, \emph{i.e.,} their Bayesian posteriors can be analytically derived and have the same functional forms as the priors (see Ref.~\cite{Melendez:2019izc}). The assumptions made above are known to impose certain limitations~\cite{bastos_ohagan}; however their validity can be assessed from several diagnostic metrics on the validation data set (see below).
	
\paragraph*{Application to $\sigma_{np}$ and $\sigma_{\gamma d}$---}We now apply the tools described above to the $np\leftrightarrow d\gamma$ cross section.
We begin by partitioning the data, \emph{i.e.,} the order-by-order $\chi$EFT results depicted in Fig.~\ref{fig:compare_to_expt}, into training and validation sets. This data set consists of 32 values of $p$ in 1~MeV increment starting form the smallest that corresponds to thermal-neutron $p(n,\gamma)d$. These values span a range of approximately 0 to 1~MeV~(2~MeV) in $E$~($E_n$), and encompass the $4~\mathrm{MeV}<p<14~\mathrm{MeV}$ interval most relevant for BBN. We use every fifth point for training and the rest for validation. We will see further below that our maximum \emph{a posteriori} (MAP) value of $\ell$ vindicates this grid and training-validation splitting. Similar results are obtained for other reasonable choices. As in Ref.~\cite{Melendez:2019izc}, we take $\Lambda=600$~MeV and $Q(p) = (p^8+m_\pi^8)/(p^7+m_\pi^7)\,,$
which give an expansion parameter of approximately 0.23 with a very weak dependence on $p$. We take $y_\mathrm{ref}(p)=y_0(p)$. The coefficient $c_5(p)$ turns out to be rather unnaturally sized, particularly at smaller values of $p$. We therefore exclude it from the statistical analysis. This leaves $c_{2,3,4}(p)$ for GP modeling. Finally, we need a value for the \emph{nugget} $\sigma_n^2$, the variance of the Gaussian white noise added to the data to stabilize matrix inversions during fitting. We find that $\sigma_n^2>10^{-8}$ is required to avoid singularities and that $\sigma_n^2>10^{-3}$ can introduce noises similar in size to our precision goal of 0.1-0.2\% on the predicted cross sections. We therefore pick $\sigma_n^2=10^{-5}$ as this value produces the best performing model on the validation set under the diagnostic criteria discussed below. 

\section{Model calibration, validation and prediction} 
\label{sec:gpmodel}
 \begin{figure*}[th]
 	\centering
 	\includegraphics[width=0.9\textwidth]{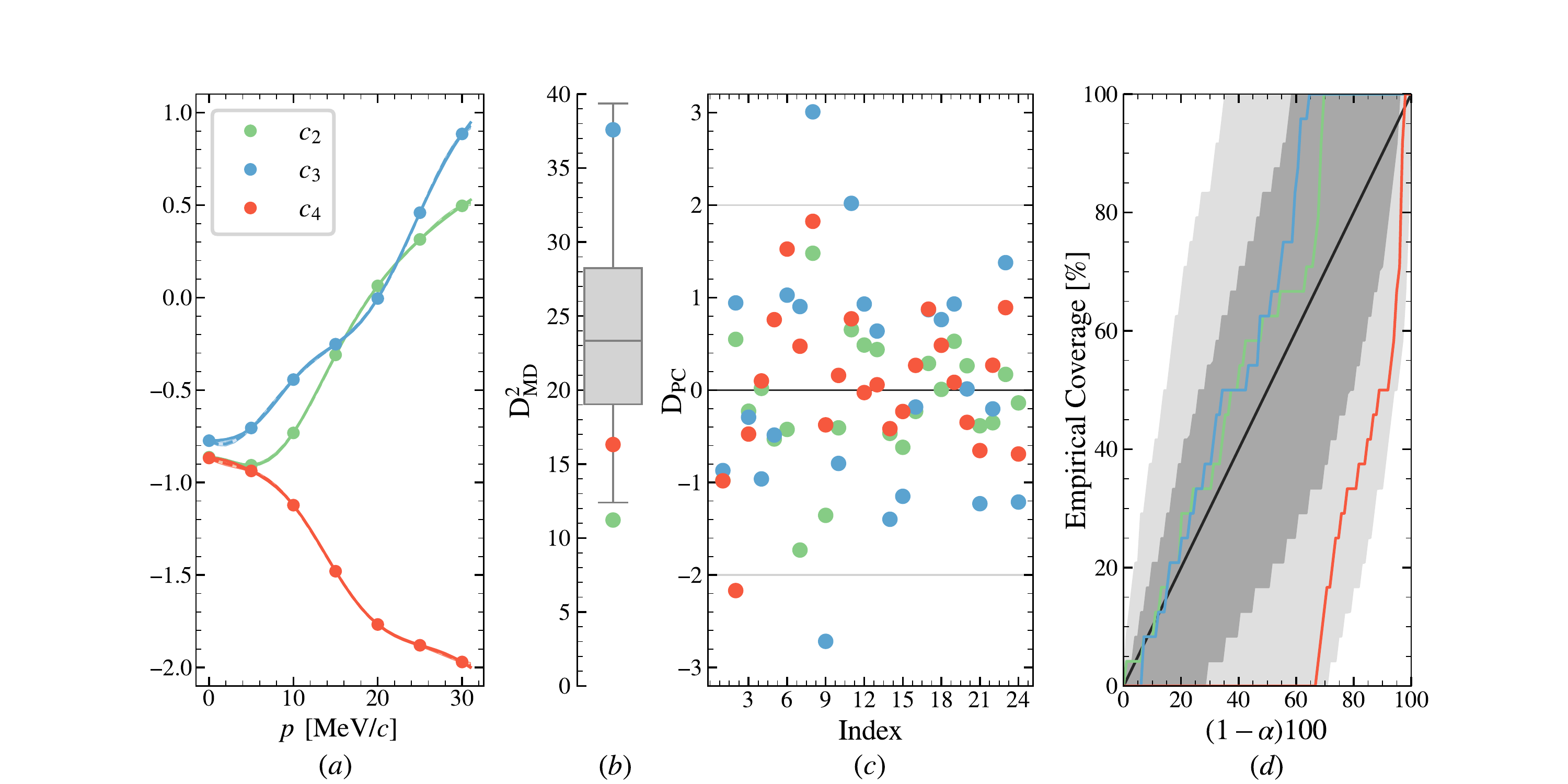}
 	\caption{GP modeling of the $\chi$EFT expansion coefficients and its diagnostics. (a) The simulators (solid lines) along with the corresponding GP emulators (dashed lines) and their $2\sigma$ intervals (bands). The training data are denoted by filled circles; 4 validation points are located uniformly between each adjacent pair of training points. (b) The Mahalanobis distances compared to the mean (interior line), $50\%$ (box) and $95\%$ (whiskers) credible intervals of the reference distribution. (c) The pivoted Cholesky diagnostics versus the index along with $95\%$ credible intervals (gray lines). (d) The credible interval diagnostics with $1\sigma$ (dark gray) and $2\sigma$ (light gray) bands estimated by sampling 1000 GP emulators.}
 	\label{fig:diagnostics}
 \end{figure*}
 
We now present the results of the GP modeling, obtained using the package {\tt gsum}~\cite{Melendez:2019izc}. Fig.~\ref{fig:diagnostics}(a) shows the coefficients $c_n(p)$ for the observable $\sigma_{np}$, along with their GP emulators. The MAP value of $\ell$ is found to be $\ell_\mathrm{MAP}=$10.4~MeV. Our choices of the grid and training-validation splitting places the training points at 5 MeV intervals. This results in 3 training data points within one correlation length, which is optimal for GP modeling. Furthermore, our data set spans approximately $3\ell_\mathrm{MAP}$ which, as a rule of thumb, is the range beyond which the data points become uncorrelated.

Fig.~\ref{fig:diagnostics}(a) provides a beautiful visual indication that the GP model has done an excellent job of emulating the actual $\chi$EFT calculations. However, detailed diagnostic checks are needed to quantitatively assess the adequacy of the model. To this end, we use the diagnostic metrics proposed by Ref.~\cite{bastos_ohagan}, and originally implemented in EFT error model by Ref.~\cite{Melendez:2019izc}. 

 Fig.~\ref{fig:diagnostics}(b) shows the squared  Mahalanobis distances, defined as 
\begin{equation}
 \mathrm{D}^2_\mathrm{MD} = (\mathbf{f}-\mathbf{m})^\mathrm{T} K^{-1} (\mathbf{f}-\mathbf{m})\,,
\end{equation}
where we have used the notation $\mathbf{f}$ for the vector of validation data points, $\mathbf{m}$ for the vector of means of the emulator at these points and $K$ for its covariance matrix. $\mathrm{D}^2_\mathrm{MD}$ is a generalization of the sum of squared residuals to the case of data points that are correlated across the independent variable. Values much larger than its reference distribution, a $\chi^2$, indicate conflict between the emulator and the simulator. Values much smaller than the reference, as we see in the case of $c_2$ to some extent, indicate that, given statistical fluctuations, there is an unusually close match between the emulator and the simulator. 

We now look at a more informative metric, the pivoted Cholesky diagnostic $\mathrm{D}_\mathrm{PC}$. For a ``correct" emulator, it returns draws from a standard Gaussian at every index, which represents the validation points arranged such that the first element is the one with the largest predictive variance, the second element is the one with the largest variance conditioned on the first element, and so on. A group of unusually large or small $\mathrm{D}_\mathrm{PC}$ values across all indices indicates a misestimated variance whereas a group of unusually large or small $\mathrm{D}_\mathrm{PC}$ values in the latter part of the sequence indicates an inappropriate correlation structure. Overall, Fig.~\ref{fig:diagnostics}(c) shows that the points are distributed as expected, \emph{e.g.,} 4 out of the 72 points lie outside the $-2<\mathrm{D}_\mathrm{PC}< 2$ range. We also notice that there is a slight indication that the variance on $c_2$ ($c_3$) might have been somewhat overestimated (underestimated) by observing the spread of the corresponding data points. 

 The credible interval diagnostic involves constructing uncertainty bands at each order and checking whether it actually encompasses the correction that enters at the next order. The claimed $(1-\alpha)100\%$ credible intervals are then plotted against the percentage of validation data points found within the interval---emulators that output credible intervals containing too few data points compared to the reference distribution are overconfident and those that contain too many are underconfident. For uncorrelated data points, the reference distribution is a binomial. For correlated data points, the reference distribution is numerically estimated by sampling a large number of emulators from the underlying process. Fig.~\ref{fig:diagnostics}(d) shows that the model is performing as expected and that it is important to account for correlations while assigning truncation errors.  
 
\begin{figure*}[ht]
	\centering
	\includegraphics[width=0.75\textwidth]{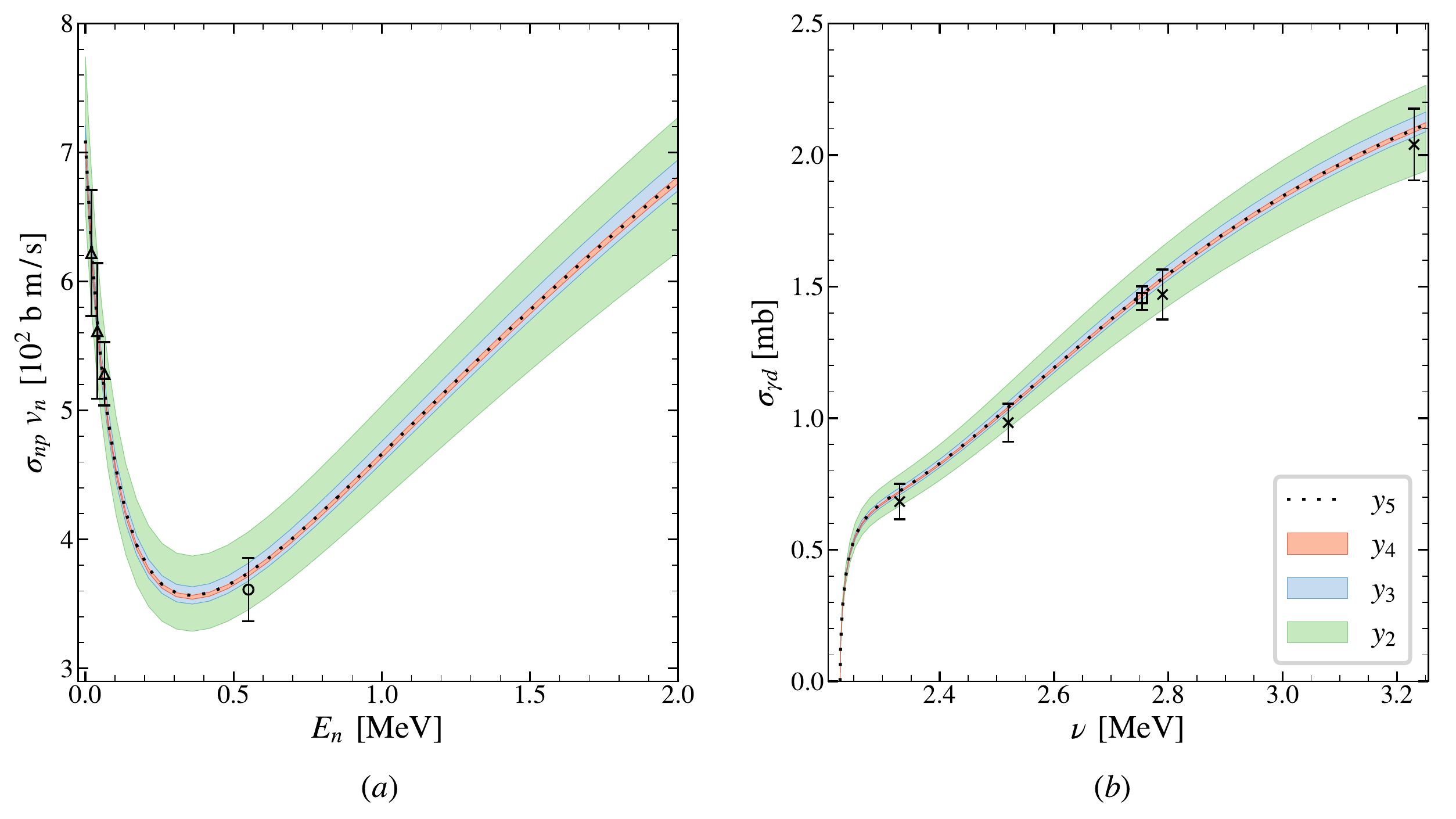}
	\caption{The $2\sigma$ truncation error bands on the $\chi$EFT predictions $y_k$ at $k=2,3,4$ along with the prediction $y_5$ and data from Fig.~\ref{fig:compare_to_expt}. $(a)$ The product of $p(n,\gamma)d$ cross section and the neutron speed versus the energy of the neutron. $(b)$ The deuteron photodissociation cross section as a function of the photon energy in the rest frame of the deuteron. }
	\label{fig:uncertainty}
    \end{figure*}

 Now that we have demonstrated that the coefficients $c_{2,3,4}$ can be appropriately described by a GP, we can now use Eq.~\eqref{eq:trunc_err_distr} to compute the truncation errors. We list the $np\rightarrow d\gamma$ cross section values at order $k=4$ along with their $1\sigma$ truncation errors at several energies in Table~\ref{tab:uncertainty}. We note that these errors are different from, and vary much more smoothly with $E$, than naive estimates~\cite{PhysRevLett.115.122301} based on multiplying the largest order-to-order shift with the appropriate power of the expansion parameter in a pointwise manner at each value of $E$. In Fig.~\ref{fig:uncertainty}(a), we plot $\sigma_{np}v_n$, for which we showed order-by-order results earlier in Fig.~\ref{fig:compare_to_expt}(a), versus $E_n$. This quantity is proportional to the reaction rate in BBN. The bands represent 2$\sigma$ truncation errors, \emph{i.e.,} 95\% Bayesian credible intervals for $\sigma_{np}v_n$. In Fig.~\ref{fig:uncertainty}(b), we show these bands for $\sigma_{\gamma d}$ and compare them with photodissociation data shown earlier in Fig.~\ref{fig:compare_to_expt}(b). Reassuringly, the experimental data as well as the highest-order theoretical calculation, $y_5$, are compatible with the truncation error estimates.

\begin{center}
\begin{table}[htbp]
\begin{tabularx}{0.75\columnwidth}{cc}
\hline
$E$ [MeV] &~~~   $\sigma_{np}$ [mb] \\
\hline
~~~$1.262500\times10^{-08}$ &~~~   $321.009 \pm 0.71496$ \\
~~~$9.607513\times10^{-03}$ &~~~   $0.32739 \pm 0.00073$ \\
~~~$3.838601\times10^{-02}$ &~~~   $0.12762 \pm 0.00029$ \\
~~~$8.633551\times10^{-02}$ &~~~   $0.06853 \pm 0.00015$ \\
~~~$1.534560\times10^{-01}$ &~~~   $0.04658 \pm 0.00010$ \\
~~~$2.397475\times10^{-01}$ &~~~   $0.03792 \pm 0.00008$ \\
~~~$3.452100\times10^{-01}$ &~~~   $0.03464 \pm 0.00007$ \\
~~~$4.698435\times10^{-01}$ &~~~   $0.03368 \pm 0.00007$ \\
~~~$6.136480\times10^{-01}$ &~~~   $0.03373 \pm 0.00007$ \\
~~~$7.766235\times10^{-01}$ &~~~   $0.03414 \pm 0.00007$ \\
~~~$9.587699\times10^{-01}$ &~~~   $0.03461 \pm 0.00007$ \\
\hline
\end{tabularx}
\caption{$\chi$EFT predictions at order $k=4$ for the $np\rightarrow d\gamma$ cross section at $np$ relative energy $E$, along with their $1\sigma$ errors from the truncation of the $\chi$EFT potential.}
\label{tab:uncertainty}
\end{table}
\end{center}

 The uncertainties quoted in Table~\ref{tab:uncertainty} and Fig.~\ref{fig:uncertainty}, which amount to about 0.2$\%$, only include truncation error in the $\chi$EFT potential. The full theory uncertainty also comprises statistical error from fitting the LECs to experimental data. The framework we have adopted allows one to also incorporate fitting uncertainties on the LECs that appear up to the EFT order we have calculated~\cite{Melendez:2019izc}. Another missing source of uncertainty, which is important in the $M1$-dominated regime, is the truncation of the current. Inclusion of the $M1$ operator at order $(Q/\Lambda)^1$ is crucial for obtaining agreement with the experimental value for threshold capture given in Table~\ref{tab:threshold}~\cite{piarulli,Phillips:2016mov}, although it introduces several new LECs and thus poses significant challenges for rigorous uncertainty analysis. The fitting strategy that uses minimal assumptions about the short-distance behavior of the current operator, among several explored by Ref.~\cite{piarulli}, is the one that constrains the LECs $d_{1,2}^V$ simultaneously to $\sigma_{np}(E=1.2625\times10^{-08}~\mathrm{MeV})$ and the isovector combination of the $A=3$ magnetic moments. However, it was found that this yields unnatural values for $d_{1,2}^V$.  This fine-tuning can be mitigated by including the theory uncertainty we calculated in this paper, as well as the experimental error, in the fit. Such a strategy for performing parameter estimation with $\chi$EFT truncation error included as a guard against overfitting was recently successfully pursued by Ref.~\cite{Wesolowski:2021cni} in the context of constraining 3N interactions from properties of $A=3,4$ nuclei. A calculation of $np\leftrightarrow d\gamma$ along these lines is a subject for future work. \newline{}

\section{Summary and outlook}
\label{sec:conclusion}

We performed the first $\chi$EFT calculations of the energy-dependent $np\leftrightarrow d\gamma$ cross section at low energies, including the range relevant to BBN, and the first Bayesian analysis of $\chi$EFT truncation error to a nuclear reaction cross section. Working with fixed one- and two-body electromagnetic current operators, we studied the convergence of this observable in the EFT expansion of the potential. By harnessing recent progress in Bayesian analysis of EFT uncertainty,
we were able to provide statistical estimates, amounting to 0.2$\%$, for the theory uncertainty that stems from truncation of the $\chi$EFT potential.

At the $\chi$EFT order up to which we work, our calculations are pure predictions as no new LECs enter that need to be fixed. 
 Inclusion of the next order in the current operator adds new parameters that are not well constrained at present and will most likely require fitting to electromagnetic observables in $A=2,3$ systems. To make predictions for this cross section with subpercentage-level precision even at threshold, we will therefore need further investigation into the nature of the current operator at short distances and calculation of the electromagnetic observables for $A=3$ nuclei with NN and 3N interactions truncated consistently. Finally, uncertainty analysis in theoretical calculations of Deuterium burning processes will be an important development because these strongly affect the primordial Deuterium abundance and there is currently some discrepancy between theory and experiments, most notably for ${}^2$H$(p,\gamma){}^3$He~\cite{Pisanti:2020efz}.

\section*{Acknowledgements}

We are grateful to Jordan Melendez, Richard J Furnstahl and Daniel R Phillips for fruitful discussions. We would also like to thank Daniel R Phillips for a critical reading of the manuscript, Evgeny Epelbaum for providing us with computer programs for chiral potentials used in this work and the BUQEYE collaboration for making the library {\tt gsum} publicly accessible. This work was supported by the Deutsche Forschungsgemeinschaft (DFG, German Research Foundation), through the 
Cluster of Excellence [Precision Physics, Fundamental Interactions, and Structure of Matter] (PRISMA$^+$ EXC 2118/1) 
within the German Excellence Strategy (Project ID 39083149).


\begin{thebibliography}{10}
	\expandafter\ifx\csname url\endcsname\relax
	\def\url#1{\texttt{#1}}\fi
	\expandafter\ifx\csname urlprefix\endcsname\relax\def\urlprefix{URL }\fi
	\expandafter\ifx\csname href\endcsname\relax
	\def\href#1#2{#2} \def\path#1{#1}\fi
	
	\bibitem{Epelbaum:2008ga}
	E.~Epelbaum, H.-W. Hammer, U.-G. Mei{\ss}ner, {Modern Theory of Nuclear
		Forces}, Rev. Mod. Phys. 81 (2009) 1773.
	\newblock \href {https://doi.org/10.1103/RevModPhys.81.1773}
	{\path{doi:10.1103/RevModPhys.81.1773}}.
	
	\bibitem{Schindler:2008fh}
	M.~R. Schindler, D.~R. Phillips, {Bayesian Methods for Parameter Estimation in
		Effective Field Theories}, Annals Phys. 324 (2009) 682, [Erratum: Annals
	Phys. 324 (2009) 2051].
	\newblock \href {https://doi.org/10.1016/j.aop.2008.09.003}
	{\path{doi:10.1016/j.aop.2008.09.003}}.
	
	\bibitem{Furnstahl:2014xsa}
	R.~J. Furnstahl, D.~R. Phillips, S.~Wesolowski, {A recipe for EFT uncertainty
		quantification in nuclear physics}, J. Phys. G 42 (2015) 034028.
	\newblock \href {https://doi.org/10.1088/0954-3899/42/3/034028}
	{\path{doi:10.1088/0954-3899/42/3/034028}}.
	
	\bibitem{Furnstahl:2015rha}
	R.~J. Furnstahl, N.~Klco, D.~R. Phillips, S.~Wesolowski, {Quantifying
		truncation errors in effective field theory}, Phys. Rev. C 92 (2015) 024005.
	\newblock \href {https://doi.org/10.1103/PhysRevC.92.024005}
	{\path{doi:10.1103/PhysRevC.92.024005}}.
	
	\bibitem{Wesolowski:2015fqa}
	S.~Wesolowski, N.~Klco, R.~J. Furnstahl, D.~R. Phillips, A.~Thapaliya,
	{Bayesian parameter estimation for effective field theories}, J. Phys. G 43
	(2016) 074001.
	\newblock \href {https://doi.org/10.1088/0954-3899/43/7/074001}
	{\path{doi:10.1088/0954-3899/43/7/074001}}.
	
	\bibitem{Zhang:2015ajn}
	X.~Zhang, K.~M. Nollett, D.~R. Phillips, {Halo effective field theory
		constrains the solar $^7$Be + p $rightarrow$  $^8$B + \ensuremath{\gamma}
		rate}, Phys. Lett. B 751 (2015) 535.
	\newblock \href {https://doi.org/10.1016/j.physletb.2015.11.005}
	{\path{doi:10.1016/j.physletb.2015.11.005}}.
	
	\bibitem{Melendez:2017phj}
	J.~A. Melendez, S.~Wesolowski, R.~J. Furnstahl, {Bayesian truncation errors in
		chiral effective field theory: nucleon-nucleon observables}, Phys. Rev. C 96
	(2017) 024003.
	\newblock \href {https://doi.org/10.1103/PhysRevC.96.024003}
	{\path{doi:10.1103/PhysRevC.96.024003}}.
	
	\bibitem{Wesolowski:2018lzj}
	S.~Wesolowski, R.~J. Furnstahl, J.~A. Melendez, D.~R. Phillips, {Exploring
		Bayesian parameter estimation for chiral effective field theory using
		nucleon\textendash{}nucleon phase shifts}, J. Phys. G 46 (2019) 045102.
	\newblock \href {https://doi.org/10.1088/1361-6471/aaf5fc}
	{\path{doi:10.1088/1361-6471/aaf5fc}}.
	
	\bibitem{Lovell:2018bao}
	A.~E. Lovell, F.~M. Nunes, {Constraining Transfer Cross Sections Using Bayes'
		Theorem}, Phys. Rev. C 97 (2018) 064612.
	\newblock \href {https://doi.org/10.1103/PhysRevC.97.064612}
	{\path{doi:10.1103/PhysRevC.97.064612}}.
	
	\bibitem{Neufcourt:2018syo}
	L.~Neufcourt, Y.~Cao, W.~Nazarewicz, F.~Viens, {Bayesian approach to
		model-based extrapolation of nuclear observables}, Phys. Rev. C 98 (2018)
	034318.
	\newblock \href {https://doi.org/10.1103/PhysRevC.98.034318}
	{\path{doi:10.1103/PhysRevC.98.034318}}.
	
	\bibitem{Zhou:2018bon}
	S.~Zhou, P.~Giulani, J.~Piekarewicz, A.~Bhattacharya, D.~Pati, {Reexamining the
		proton-radius problem using constrained Gaussian processes}, Phys. Rev. C 99
	(2019) 055202.
	\newblock \href {https://doi.org/10.1103/PhysRevC.99.055202}
	{\path{doi:10.1103/PhysRevC.99.055202}}.
	
	\bibitem{King:2019sax}
	G.~B. King, A.~E. Lovell, L.~Neufcourt, F.~M. Nunes, {Direct comparison between
		Bayesian and frequentist uncertainty quantification for nuclear reactions},
	Phys. Rev. Lett. 122 (2019) 232502.
	\newblock \href {https://doi.org/10.1103/PhysRevLett.122.232502}
	{\path{doi:10.1103/PhysRevLett.122.232502}}.
	
	\bibitem{Melendez:2019izc}
	J.~A. Melendez, R.~J. Furnstahl, D.~R. Phillips, M.~T. Pratola, S.~Wesolowski,
	{Quantifying Correlated Truncation Errors in Effective Field Theory}, Phys.
	Rev. C 100 (2019) 044001.
	\newblock \href {https://doi.org/10.1103/PhysRevC.100.044001}
	{\path{doi:10.1103/PhysRevC.100.044001}}.
	
	\bibitem{Premarathna:2019tup}
	P.~Premarathna, G.~Rupak, {Bayesian analysis of capture reactions
		$^3\mathrm{He}(\alpha,\gamma)^7\mathrm{Be}$ and
		$^3\mathrm{H}(\alpha,\gamma)^7\mathrm{Li}$}, Eur. Phys. J. A 56 (2020) 166.
	\newblock \href {https://doi.org/10.1140/epja/s10050-020-00113-z}
	{\path{doi:10.1140/epja/s10050-020-00113-z}}.
	
	\bibitem{Zhang:2019odg}
	X.~Zhang, K.~M. Nollett, D.~R. Phillips, {$S$-factor and scattering-parameter
		extractions from ${}^{3}\mathrm{He} +{}^{4}\mathrm{He} \rightarrow
		{}^{7}\mathrm{Be} + \gamma$}, J. Phys. G 47 (2020) 054002.
	\newblock \href {https://doi.org/10.1088/1361-6471/ab6a71}
	{\path{doi:10.1088/1361-6471/ab6a71}}.
	
	\bibitem{Epelbaum:2019zqc}
	E.~Epelbaum, et~al., {Towards high-order calculations of three-nucleon
		scattering in chiral effective field theory}, Eur. Phys. J. A 56 (2020) 92.
	\newblock \href {https://doi.org/10.1140/epja/s10050-020-00102-2}
	{\path{doi:10.1140/epja/s10050-020-00102-2}}.
	
	\bibitem{Phillips:2020dmw}
	D.~R. Phillips, et~al., {Get on the BAND Wagon: A Bayesian Framework for
		Quantifying Model Uncertainties in Nuclear Dynamics}, J. Phys. G 48 (2021)
	072001.
	\newblock \href {https://doi.org/10.1088/1361-6471/abf1df}
	{\path{doi:10.1088/1361-6471/abf1df}}.
	
	\bibitem{Volkotrub:2020lsr}
	Y.~Volkotrub, J.~Golak, R.~Skibi\'nski, K.~Topolnicki, H.~Wita\l{}a,
	E.~Epelbaum, H.~Krebs, P.~Reinert, {Uncertainty of three-nucleon continuum
		observables arising from uncertainties of two-nucleon potential parameters},
	J. Phys. G 47 (2020) 104001.
	\newblock \href {https://doi.org/10.1088/1361-6471/ab9b07}
	{\path{doi:10.1088/1361-6471/ab9b07}}.
	
	\bibitem{Reinert:2020mcu}
	P.~Reinert, H.~Krebs, E.~Epelbaum, {Precision determination of pion-nucleon
		coupling constants using effective field theory}, Phys. Rev. Lett. 126 (2021)
	092501.
	\newblock \href {https://doi.org/10.1103/PhysRevLett.126.092501}
	{\path{doi:10.1103/PhysRevLett.126.092501}}.
	
	\bibitem{Wesolowski:2021cni}
	S.~Wesolowski, I.~Svensson, A.~Ekstr\"om, C.~Forss\'en, R.~J. Furnstahl, J.~A.
	Melendez, D.~R. Phillips, {Rigorous constraints on three-nucleon forces in chiral effective field theory from fast and accurate calculations of few-body observables}, Phys. Rev. C 104 (2021)
	064001. \newblock \href {https://doi.org/10.1103/PhysRevC.104.064001} {\path{doi:10.1103/PhysRevC.104.064001}}.
	
	\bibitem{Rijneveen:2020qbc}
	J.~Rijneveen, N.~Rijneveen, H.~Krebs, A.~M. Gasparyan, E.~Epelbaum, {Radiative
		pion photoproduction in covariant chiral perturbation theory}, Phys. Rev. C
	103 (2021) 045203.
	\newblock \href {https://doi.org/10.1103/PhysRevC.103.045203}
	{\path{doi:10.1103/PhysRevC.103.045203}}.
	
	\bibitem{Maris:2020qne}
	P.~Maris, et~al., {Light nuclei with semilocal momentum-space regularized
		chiral interactions up to third order}, Phys. Rev. C 103 (2021) 054001.
	\newblock \href {https://doi.org/10.1103/PhysRevC.103.054001}
	{\path{doi:10.1103/PhysRevC.103.054001}}.
	
	\bibitem{Wellenhofer:2021eis}
	C.~Wellenhofer, C.~Drischler, A.~Schwenk, {Effective field theory for dilute
		Fermi systems at fourth order}, Phys. Rev. C 104 (2021) 014003.
	\newblock \href {https://doi.org/10.1103/PhysRevC.104.014003}
	{\path{doi:10.1103/PhysRevC.104.014003}}.
	
	\bibitem{Connell:2021qcd}
	M.~A. Connell, I.~Billig, D.~R. Phillips, {Does Bayesian model averaging
		improve polynomial extrapolations? Two toy problems as tests}, J. Phys. G 48
	(2021) 104001.
	\newblock \href {https://doi.org/10.1088/1361-6471/ac215a}
	{\path{doi:10.1088/1361-6471/ac215a}}.
	
	\bibitem{Rijneveen:2021bfw}
	N.~Rijneveen, A.~M. Gasparyan, H.~Krebs, E.~Epelbaum, {Pion photoproduction in
		chiral perturbation theory with explicit treatment of the $\Delta(1232)$
		resonance }\href {http://arxiv.org/abs/2108.01619} {\path{arXiv:2108.01619}}.
	
	\bibitem{Mossa:2020gjc}
	V.~Mossa, et~al., {The baryon density of the Universe from an improved rate of
		deuterium burning}, Nature 587 (2020) 210.
	\newblock \href {https://doi.org/10.1038/s41586-020-2878-4}
	{\path{doi:10.1038/s41586-020-2878-4}}.
	
	\bibitem{Serpico_2004}
	P.~D. Serpico, S.~Esposito, F.~Iocco, G.~Mangano, G.~Miele, O.~Pisanti, Nuclear
	reaction network for primordial nucleosynthesis: a detailed analysis of
	rates, uncertainties and light nuclei yields, Journal of Cosmology and
	Astroparticle Physics 12 (2004) 010.
	\newblock \href {https://doi.org/10.1088/1475-7516/2004/12/010}
	{\path{doi:10.1088/1475-7516/2004/12/010}}.
	
	\bibitem{Pisanti:2007hk}
	O.~Pisanti, A.~Cirillo, S.~Esposito, F.~Iocco, G.~Mangano, G.~Miele, P.~D.
	Serpico, {PArthENoPE: Public Algorithm Evaluating the Nucleosynthesis of
		Primordial Elements}, Comput. Phys. Commun. 178 (2008) 956.
	\newblock \href {https://doi.org/10.1016/j.cpc.2008.02.015}
	{\path{doi:10.1016/j.cpc.2008.02.015}}.
	
	\bibitem{Coc:2019rza}
	A.~Coc, C.~Pitrou, J.-P. Uzan, E.~Vangioni, {A Public Code for Precision Big
		Bang Nucleosynthesis with Improved Helium-4 Predictions}, Springer Proc.
	Phys. 219 (2019) 327.
	\newblock \href {https://doi.org/10.1007/978-3-030-13876-9_57}
	{\path{doi:10.1007/978-3-030-13876-9_57}}.
	
	\bibitem{Pitrou:2018cgg}
	C.~Pitrou, A.~Coc, J.-P. Uzan, E.~Vangioni, {Precision big bang nucleosynthesis
		with improved Helium-4 predictions}, Phys. Rept. 754 (2018) 1.
	\newblock \href {https://doi.org/10.1016/j.physrep.2018.04.005}
	{\path{doi:10.1016/j.physrep.2018.04.005}}.
	
	\bibitem{Planck:2018vyg}
	N.~Aghanim, et~al., {Planck 2018 results. VI. Cosmological parameters}, Astron.
	Astrophys. 641 (2020) A6, [Erratum: Astron.Astrophys. 652, C4 (2021)].
	\newblock \href {https://doi.org/10.1051/0004-6361/201833910}
	{\path{doi:10.1051/0004-6361/201833910}}.
	
	\bibitem{Cooke:2017cwo}
	R.~J. Cooke, M.~Pettini, C.~C. Steidel, {One Percent Determination of the
		Primordial Deuterium Abundance}, Astrophys. J. 855 (2018) 102.
	\newblock \href {https://doi.org/10.3847/1538-4357/aaab53}
	{\path{doi:10.3847/1538-4357/aaab53}}.
	
	\bibitem{Arenhovel:1990yg}
	H.~Arenhovel, M.~Sanzone, {Photodisintegration of the deuteron: A Review of
		theory and experiment}, Few Body Syst. Suppl. 3 (1991) 1.
	
	\bibitem{Carlson:1997qn}
	J.~Carlson, R.~Schiavilla, {Structure and Dynamics of Few Nucleon Systems},
	Rev. Mod. Phys. 70 (1998) 743.
	\newblock \href {https://doi.org/10.1103/RevModPhys.70.743}
	{\path{doi:10.1103/RevModPhys.70.743}}.
	
	\bibitem{Marcucci:2004sq}
	L.~E. Marcucci, K.~M. Nollett, R.~Schiavilla, R.~B. Wiringa, {Modern theories
		of low-energy astrophysical reactions}, Nucl. Phys. A 777 (2006) 111.
	\newblock \href {https://doi.org/10.1016/j.nuclphysa.2004.09.008}
	{\path{doi:10.1016/j.nuclphysa.2004.09.008}}.
	
	\bibitem{chen-savage}
	J.-W. Chen, M.~J. Savage, {$\ensuremath{np\rightarrow d\gamma}$ for big bang
		nucleosynthesis}, Phys. Rev. C 60 (1999) 065205.
	\newblock \href {https://doi.org/10.1103/PhysRevC.60.065205}
	{\path{doi:10.1103/PhysRevC.60.065205}}.
	
	\bibitem{rupak}
	G.~Rupak, Precision calculation of $\ensuremath{np\rightarrow d\gamma}$ cross
	section for big-bang nucleosynthesis, Nuclear Physics A 678 (2000) 405.
	\newblock \href {https://doi.org/https://doi.org/10.1016/S0375-9474(00)00323-7}
	{\path{doi:https://doi.org/10.1016/S0375-9474(00)00323-7}}.
	
	\bibitem{ando}
	S.~Ando, R.~H. Cyburt, S.~W. Hong, C.~H. Hyun, Radiative neutron capture on a
	proton at big-bang nucleosynthesis energies, Phys. Rev. C 74 (2006) 025809.
	\newblock \href {https://doi.org/10.1103/PhysRevC.74.025809}
	{\path{doi:10.1103/PhysRevC.74.025809}}.
	
	\bibitem{cokinos-melkonian}
	D.~Cokinos, E.~Melkonian, Measurement of the 2200 m/sec neutron-proton capture
	cross section, Phys. Rev. C 15 (1977) 1636.
	\newblock \href {https://doi.org/10.1103/PhysRevC.15.1636}
	{\path{doi:10.1103/PhysRevC.15.1636}}.
	
	\bibitem{Beane:2015yha}
	S.~R. Beane, E.~Chang, W.~Detmold, K.~Orginos, A.~Parre\~no, M.~J. Savage,
	B.~C. Tiburzi, {Ab initio Calculation of the
		$np\rightarrow{}d\ensuremath{\gamma}$ Radiative Capture Process}, Phys.
	Rev. Lett. 115 (2015) 132001.
	\newblock \href {https://doi.org/10.1103/PhysRevLett.115.132001}
	{\path{doi:10.1103/PhysRevLett.115.132001}}.
	
	\bibitem{piarulli}
	M.~Piarulli, L.~Girlanda, L.~E. Marcucci, S.~Pastore, R.~Schiavilla,
	M.~Viviani, Electromagnetic structure of $\ensuremath{A=2}$ and 3 nuclei in
	chiral effective field theory, Phys. Rev. C 87 (2013) 014006.
	\newblock \href {https://doi.org/10.1103/PhysRevC.87.014006}
	{\path{doi:10.1103/PhysRevC.87.014006}}.
	
	\bibitem{song-lazauskas-park}
	Y.-H. Song, R.~Lazauskas, T.-S. Park, Up to $\mathrm{N}^{3}\mathrm{LO}$
	heavy-baryon chiral perturbation theory calculation for the $\ensuremath{M1}$
	properties of three-nucleon systems, Phys. Rev. C 79 (2009) 064002.
	\newblock \href {https://doi.org/10.1103/PhysRevC.79.064002}
	{\path{doi:10.1103/PhysRevC.79.064002}}.
	
	\bibitem{PhysRevLett.82.4176}
	S.~Burles, K.~M. Nollett, J.~W. Truran, M.~S. Turner, Sharpening the
	predictions of big-bang nucleosynthesis, Phys. Rev. Lett. 82 (1999) 4176.
	\newblock \href {https://doi.org/10.1103/PhysRevLett.82.4176}
	{\path{doi:10.1103/PhysRevLett.82.4176}}.
	
	\bibitem{PhysRevC.102.054315}
	C.~Drischler, J.~A. Melendez, R.~J. Furnstahl, D.~R. Phillips, Quantifying
	uncertainties and correlations in the nuclear-matter equation of state, Phys.
	Rev. C 102 (2020) 054315.
	\newblock \href {https://doi.org/10.1103/PhysRevC.102.054315}
	{\path{doi:10.1103/PhysRevC.102.054315}}.
	
	\bibitem{PhysRevLett.125.202702}
	C.~Drischler, R.~J. Furnstahl, J.~A. Melendez, D.~R. Phillips, How well do we
	know the neutron-matter equation of state at the densities inside neutron
	stars? a bayesian approach with correlated uncertainties, Phys. Rev. Lett.
	125 (2020) 202702.
	\newblock \href {https://doi.org/10.1103/PhysRevLett.125.202702}
	{\path{doi:10.1103/PhysRevLett.125.202702}}.
	
	\bibitem{Reinert:2017usi}
	P.~Reinert, H.~Krebs, E.~Epelbaum, {Semilocal momentum-space regularized chiral
		two-nucleon potentials up to fifth order}, Eur. Phys. J. A 54 (2018) 86.
	\newblock \href {https://doi.org/10.1140/epja/i2018-12516-4}
	{\path{doi:10.1140/epja/i2018-12516-4}}.
	
	\bibitem{Park:1995pn}
	T.-S. Park, D.-P. Min, M.~Rho, {Chiral Lagrangian approach to exchange vector
		currents in nuclei}, Nucl. Phys. A596 (1996) 515--552.
	\newblock \href {http://arxiv.org/abs/nucl-th/9505017}
	{\path{arXiv:nucl-th/9505017}}, \href
	{https://doi.org/10.1016/0375-9474(95)00406-8}
	{\path{doi:10.1016/0375-9474(95)00406-8}}.
	
	\bibitem{Pastore:2008ui}
	S.~Pastore, R.~Schiavilla, J.~L. Goity, {Electromagnetic two-body currents of
		one- and two-pion range}, Phys. Rev. C78 (2008) 064002.
	\newblock \href {http://arxiv.org/abs/0810.1941} {\path{arXiv:0810.1941}},
	\href {https://doi.org/10.1103/PhysRevC.78.064002}
	{\path{doi:10.1103/PhysRevC.78.064002}}.
	
	\bibitem{Kolling:2009iq}
	S.~Kolling, E.~Epelbaum, H.~Krebs, U.-G. Mei{\ss}ner, {Two-pion exchange
		electromagnetic current in chiral effective field theory using the method of
		unitary transformation}, Phys. Rev. C80 (2009) 045502.
	\newblock \href {https://doi.org/10.1103/PhysRevC.80.045502}
	{\path{doi:10.1103/PhysRevC.80.045502}}.
	
	\bibitem{Bijaya_neutrino}
	B.~Acharya, S.~Bacca, Neutrino-deuteron scattering: Uncertainty quantification
	and new ${L}_{1,A}$ constraints, Phys. Rev. C 101 (2020) 015505.
	\newblock \href {https://doi.org/10.1103/PhysRevC.101.015505}
	{\path{doi:10.1103/PhysRevC.101.015505}}.
	
	\bibitem{chi-disp-mu-d}
	B.~Acharya, V.~Lensky, S.~Bacca, M.~Gorchtein, M.~Vanderhaeghen, Dispersive
	evaluation of the lamb shift in muonic deuterium from chiral effective field
	theory, Phys. Rev. C 103 (2021) 024001.
	\newblock \href {https://doi.org/10.1103/PhysRevC.103.024001}
	{\path{doi:10.1103/PhysRevC.103.024001}}.

	
	\bibitem{suzuki}
	T.~Suzuki, Y.~Nagai, T.~Shima, T.~Kikuchi, H.~Sato, T.~Kii, M.~Igashira, First
	measurement of a $\ensuremath{p( n, \gamma )d}$ reaction cross section
	between 10 and 80 kev, The Astrophysical Journal 439 (1995) L59.
	
	\bibitem{nagai}
	Y.~Nagai, T.~S. Suzuki, T.~Kikuchi, T.~Shima, T.~Kii, H.~Sato, M.~Igashira,
	Measurement of ${}^{1}\mathrm{H}(n,\ensuremath{\gamma}{)}^{2}\mathrm{H}$
	reaction cross section at a comparable $\ensuremath{M1/E1}$ strength, Phys.
	Rev. C 56 (1997) 3173.
	\newblock \href {https://doi.org/10.1103/PhysRevC.56.3173}
	{\path{doi:10.1103/PhysRevC.56.3173}}.
	
	\bibitem{hara}
	K.~Y. Hara, H.~Utsunomiya, S.~Goko, H.~Akimune, T.~Yamagata, M.~Ohta,
	H.~Toyokawa, K.~Kudo, A.~Uritani, Y.~Shibata, Y.-W. Lui, H.~Ohgaki,
	Photodisintegration of deuterium and big bang nucleosynthesis, Phys. Rev. D
	68 (2003) 072001.
	\newblock \href {https://doi.org/10.1103/PhysRevD.68.072001}
	{\path{doi:10.1103/PhysRevD.68.072001}}.
	
	\bibitem{moreh}
	R.~Moreh, T.~J. Kennett, W.~V. Prestwich,
	$^{2}\mathrm{H}$(\ensuremath{\gamma},n) absolute cross section at 2754 kev,
	Phys. Rev. C 39 (1989) 1247.
	\newblock \href {https://doi.org/10.1103/PhysRevC.39.1247}
	{\path{doi:10.1103/PhysRevC.39.1247}}.
	
	\bibitem{tornow}
	W.~Tornow, N.~Czakon, C.~Howell, A.~Hutcheson, J.~Kelley, V.~Litvinenko,
	S.~Mikhailov, I.~Pinayev, G.~Weisel, H.~Witała, Low-energy
	photodisintegration of the deuteron and big-bang nucleosynthesis, Physics
	Letters B 574 (2003) 8.
	\newblock \href
	{https://doi.org/https://doi.org/10.1016/j.physletb.2003.08.078}
	{\path{doi:https://doi.org/10.1016/j.physletb.2003.08.078}}.
	
	\bibitem{bastos_ohagan}
	L.~S. Bastos, A.~O’Hagan, Diagnostics for gaussian process emulators,
	Technometrics 51 (2009) 425.
	\newblock \href {https://doi.org/10.1198/TECH.2009.08019}
	{\path{doi:10.1198/TECH.2009.08019}}.

	
	\bibitem{PhysRevLett.115.122301}
	E.~Epelbaum, H.~Krebs, U.-G.~Mei{\ss}ner, {Precision Nucleon-Nucleon Potential at Fifth Order in the Chiral Expansion}, Phys. Rev. Lett. 115 (2015) 122301.
	\newblock \href {https://doi.org/10.1103/PhysRevLett.115.122301}
	{\path{doi:10.1103/PhysRevLett.115.122301}}.
	
	
	\bibitem{Phillips:2016mov}
	D.~R. Phillips, {Electromagnetic Structure of Two- and Three-Nucleon Systems:
		An Effective Field Theory Description}, Ann. Rev. Nucl. Part. Sci. 66 (2016)
	421--447.
	\newblock \href {https://doi.org/10.1146/annurev-nucl-102014-022321}
	{\path{doi:10.1146/annurev-nucl-102014-022321}}.
	
	\bibitem{Pisanti:2020efz}
	O.~Pisanti, G.~Mangano, G.~Miele, P.~Mazzella, {Primordial Deuterium after
		LUNA: concordances and error budget}, JCAP 04 (2021) 020.
	\newblock \href {https://doi.org/10.1088/1475-7516/2021/04/020}
	{\path{doi:10.1088/1475-7516/2021/04/020}}.
	
	\bibitem{Borah:2020gte}
	K.~Borah, R.~J. Hill, G.~Lee, O.~Tomalak, {Parametrization and applications of
		the low-$Q^2$ nucleon vector form factors}, Phys. Rev. D 102 (2020) 074012.
	\newblock \href {https://doi.org/10.1103/PhysRevD.102.074012}
	{\path{doi:10.1103/PhysRevD.102.074012}}.

	\bibitem{bibnote1}
	{To include nucleon-structure effects, we adopt a recent model-independent parametrization of the nucleon form factors~\cite{Borah:2020gte} in lieu of summing the corresponding diagrams in $\chi$EFT.}
	
\end{thebibliography}
\end{document}